\documentclass[twocolumn,showpacs,preprintnumbers,amsmath,amssymb]{revtex4}

\usepackage{graphicx}
\usepackage{dcolumn}
\usepackage{bm}

\def\id{{\rm d}}

\def\be{\begin{equation}}
\def\ee{\end{equation}}
\def\bea{\begin{eqnarray}}
\def\eea{\end{eqnarray}}

\begin{document}

\preprint{nakano:JBA-qubit}

\title{Quantum Time-evolution in  Qubit Readout Process\\ with a Josephson Bifurcation Amplifier}

\author{Hayato Nakano}
\email{nakano@will.brl.ntt.co.jp}
\author{Shiro Saito}%
\author{Kouichi Semba}
\affiliation{%
NTT Basic Research Laboratories, NTT Corporation\\
Atsugi-shi, Kanagawa 243-0198, Japan
}%
\author{Hideaki Takayanagi}
\affiliation{
International Center for Materials Nanoarchitectonics, NIMS\\
Tsukuba, Ibaraki 305-0003, Japan
}%

\date{\today}

\begin{abstract}

We analyzed the Josephson bifurcation amplifier (JBA) readout process 
of a superconducting qubit quantum mechanically. This was achieved by 
employing numerical analyses of the dynamics of the density operator
of a driven nonlinear oscillator and a qubit coupled system during the
measurement process. 
In purely quantum cases, the wavefunction of the JBA is
trapped in a quasienergy-state, and  bifurcation is
impossible. Introducing decoherence enables us to reproduce the
bifurcation with a finite hysteresis. 
Moreover,  we discuss in detail the dynamics involved when a qubit  is initially in 
a superposition state.  We have observed  the qubit-probe (JBA) entangled state and 
it is divided into two separable states at the moment of the JBA transition begins.
This corresponds to ``projection''.
To readout the measurement result, however, we must wait until the two JBA states are  
macroscopically well separated. The waiting time is determined
by the strength of the decoherence in the JBA.

\end{abstract}

\maketitle



\begin{figure}[h]
\includegraphics[width=5.3cm]{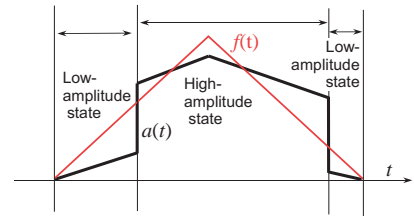}
\caption{\label{fig:schem} Hysteretic behavior of the oscillation amplitude $a(t)$ of a JBA as a function of the driving amplitude $f(t)$ (schematic). }
\end{figure}

The readout of superconducting qubit states with the Josephson bifurcation
amplifier (JBA) technique provides non-destructive and high visibility readout. 
Therefore, now it is widely and successfully used in actual experiments  \cite{Sid1}.
Mathematically, a JBA  
is described as a driven Duffing oscillator \cite{Non}. 
It enhances a small difference in  operation conditions by utilizing the bifurcation phenomenon.
Under an appropriate driving force, a {\em classical} nonlinear oscillator becomes  bistable \cite{Non}. One stable state has a small amplitude (low-amplitude state), and the other has a larger
amplitude (high-amplitude state). The critical driving force  $f_{\rm c}$ or the critical detuning $\delta_{\rm c}$ for the transition between these two states is very sensitive 
to  small changes in  the operational parameters of the oscillator. For example, when we increase or decrease the driving force continuously,
the amplitude of the oscillation behaves hysteretically as shown in Fig. \ref{fig:schem}.
When using a JBA as a qubit state readout probe, the JBA detects a small change depending on the qubit state.
However, the quantum-mechanical behavior of the JBA readout process has not been  established theoretically. 
This is because 
the bifurcation phenomenon can be
discussed  only for classical oscillators,
and is impossible from the view point of pure quantum mechanics for an isolated
system \cite{Dykman}. 
A classical analysis  gives  no information  on entanglement
between the qubit and the probe (JBA) or the decoherence in the composite
system, although all the quantum properties (projection,
measurement backaction, etc.)  in the readout are contained in such information.
A quantum mechanical analysis is indispensable if we are to 
understand the readout process. 

In this letter, we analyze the quantum-mechanical time evolution of a JBA, and clarify how a  bifurcation 
appears in an actual situation. Moreover, we invesitgate what happens during the  process of the qubit state readout 
with a JBA by analyzing dynamics of the qubit-JBA composite system.

%
In a highly quantum-mechanical JBA case, tunneling  
between classically stable states destroys the criticality
in a classical oscillator. This type of phenomenon has been precisely discussed in  \cite{Dykman} and 
in references therein.  
that the charging energy of the JBA ( $\sim 2e^2/C$, where $C$ is the effective capacitance in the JBA circuit ) is comparable to the energy barrier
($\sim $ the nonlinearity introduced below) between two stable states, and decoherecne is negligibly small. %
Howevr, actual JBA measurements are made with more classical conditions.
Rigo {\it et al.}  \cite{Rigo} investigated such an oscillator with 
a semi-classical trajectory analysis. In order to obtain quantum information more directly, here, 
we analyze the time evolution of a JBA and a qubit
during the readout process.



A JBA  can be modeled as an anhormonic oscillator in a rotating frame approximation with a Hamiltonian;
\begin{equation}\label{eq:HRF} 
H_{\rm J}=(\Omega -\omega )n_a + \alpha{n_a}^2-\frac1{2}f(a^\dagger +a)
\end{equation}
where, $a^\dagger$($a$) is the creation (annihilation) operator of the Josephson plasma oscillation. 
 $n_a =a^\dagger a$, and $\Omega$ is the linear resonant
frequency of the JBA oscillator.
$\omega$ is the driving frequency, which is  slightly smaller than $\Omega$ 
by the detuning $\delta\equiv \Omega-\omega$.
$f$ is the driving strength, and $\alpha (>0)$ is the nonlinearity.
In a classical approximation, 
this model shows the bifurcation in an appropriate parameter region.
However, for a quantum-mechanical junction with $[a,a^\dagger]=1$,  
the transition from $|G\rangle_{\rm J}$ (low-amplitude state) to $|E\rangle_{\rm J}$ (high-amplitude state)
or, from $|E\rangle_{\rm J}$  to $|G\rangle_{\rm J}$  is impossible. 



The quasienergy-states (eigenstates of the Hamiltonian Eq. (\ref{eq:HRF}) in the rotating approximation) 
are easily calculated and it is 
found that eigenstates never cross when the driving strength $f$ is changed adiabatically.
This means that if the JBA is initially in the ground state without driving, 
it never moves to the high-amplitude resonant state
even if we increase the driving field 
because 
the JBA state only moves along the initial quasienergy-state and
never jumps to the  quasienergy-state which the high-amplitude resonant state belongs.


Therefore, we expect that when a transition between quasienergy-states is caused by perturbation 
from  outside the system
the bifurcation phenomenon is reproduced. This is the case when decoherence is introduced into the present model.
Here, we only take into account of  the decoherence caused by a bath coupled to the JBA because  decoherence 
that directly attacks the qubit
is not limited to the readout process.
Even for this model, indirect decoherence via the JBA occurs in the qubit.

For example, we introduce linear loss in the oscillator (JBA). The time evolution of the system (qubit-JBA) is governed by a
Liouville equation:
\begin{equation}
\frac{\id \rho}{\id t}=\frac1{i}[\rho,H]+\frac{\Gamma}{2} (2 a \rho a^\dagger -a^\dagger a \rho -\rho a^\dagger a ),
\end{equation}
where $\rho$ is the density operator of the system, and $\Gamma$ is the relaxation rate due to the linear loss in the JBA. The $Q$-value is given by $\Omega/\Gamma$.

\begin{figure}[h]
\includegraphics[width=5.3cm]{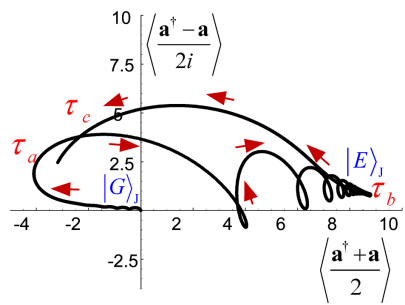}
\caption{\label{fig:Traj} (Color online)  The trajectory of the quantum expectation value $(\langle \frac{a^\dagger + a }{2}\rangle, \langle \frac{a^\dagger - a }{2i}\rangle)$ of the oscillator (complex) amplitude 
when the driving force $f$ is operated as shown in Fig. \ref{fig:schem}. 
The starting point is $|G\rangle_J$ without driving, and the right convergence point is  $|E\rangle_J$. }
\end{figure}

\begin{figure}[h]
\includegraphics[width=6.2cm]{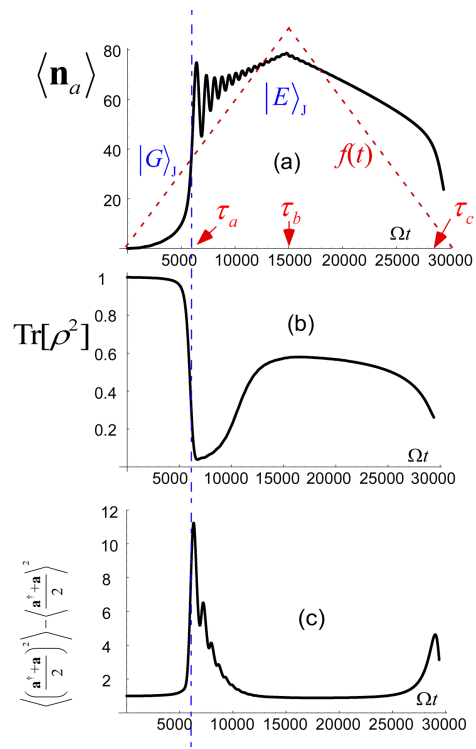}
\caption{\label{fig:3C} (Color online) Time variations of some quantum expectation values 
when the driving force $f$ is  operated as shown in Fig. \ref{fig:schem}. 
(a) Number of  bosons excited in JBA. (b) Purity of the JBA. (c) Fluctuation in JBA amplitude. }
\end{figure}

First, we show a numerical example of  JBA dynamics {\em without a qubit} in Figs. \ref{fig:Traj} and \ref{fig:3C}.
Here, the parameters used  are $\delta=0.007\Omega$, $\alpha=8\times 10^{-5}\Omega$, $Q=2500$, and $f$ is operated as
$0\rightarrow 0.025\Omega \rightarrow 0$. These parameters are similar to those used in actual experiments  \cite{Sid1}. However,
$\delta$ and $\Gamma$ are a factor of  $10^{-2}$ times smaller than real cases in order to emphasize quantumness 
and discuss the influences of decoherence.  Even if we use different parameters we obtain qualitatively same behaviors
for a JBA with similar $\delta/\Gamma$ ratio value. 

Figure \ref{fig:3C}(a) approximately corresponds to the square of the JBA amplitude shown in Fig. \ref{fig:schem}. 
So, we can see that our calculation with decoherence reproduces the bifurcation phenomenon well. 
We find that the critical driving $f_{\rm c}$ is approximately $0.011\Omega$. Once the driving exceeds this $f_{\rm c}$,
the behaviors of the JBA is the same not depending on the maximum ($f=0.025 \Omega$ for Fig. \ref{fig:3C}) driving
strength. 
Moreover, our calculation provides a lot of quantum-mechanical information about the JBA transition.
Figure \ref{fig:3C}(b) shows the time variation of the purity of the JBA state. Increasing the driving force $f$, we found that the purity  decreases abruptly ($t=\tau_{\rm a}$). This corresponds to the beginning of the transition
from  $|G\rangle_J$ to $|E\rangle_J$  of the JBA. This is a manifestation of the fact that the transition needs an 
intense emission/absorption of energy to/from an external energy bath. This energy transfer is incoherent.
After the rapid decrease, the purity recovers to some extent and the JBA approaches to the classically stable state $|E\rangle_J$ ($t=\tau_{\rm b}$). 
Since $|E\rangle_J$ is a meta-stable state (stationary point of the classical Hamiltonian ),
dragging JBA into the state by decoherence (linear loss) leads to the recovery of the purity. 
However, the purity does not reach unity because it is not a true ground state.
The fluctuation in the JBA amplitude is plotted in Fig. \ref{fig:3C}(c). 
We can see a divergence of the fluctuation at the moment of the rapid decrease in the purity ($t=\tau_{\rm b}$). 
This suggests that this JBA transition between $|G\rangle_J$  and $|E\rangle_J$ is one of  a 
phase transitions in  bosonic systems with many degrees of freedom. 

Now we discuss the criterion of decoherence that determines whether a bifurcation is observed or not.
From the above analyses we know that there is no critical value.
When  the decoherence is very small ($\Gamma <  \delta$),  the speed of the transfer
from $|G\rangle_{\rm J}$ to $|E\rangle_{\rm J}$  becomes
exponentially slower
as (schematically) $\exp[- \eta \delta/\Gamma]$, 
 where  $\eta$ is a numerical factor of the order of unity. 


Information about the qubit state is transferred to the probe (JBA) through the formation of an 
entanglement between the qubit and the probe. What we actually observe is the macroscopic state
of the JBA, and merely {\em postulate} the qubit state.  
Therefore, the process by which the entanglement is formed and split into
separable states due to decoherence (``projection'') is very important for understanding the readout process \cite{Mak}.


The qubit-JBA composite system is approximately expressed by the Hamiltonian
\begin{equation}\label{Htot}
H=H_{\rm J}+k \sigma_z n_a +H_{\rm q}, \ \ \ \ H_{\rm q}=\frac1{2}(\varepsilon \sigma_z +\Delta \sigma_x )
\end{equation}
where $H_{\rm q}$ is the Pauli operator representation of the qubit. $k$ is the interaction constant between
the qubit and the JBA. 
The qubit state ($\sigma_z$)  slightly changes
the effective detuning  $\delta +k \sigma_z$, resulting in a change in  the critical value $f_{\rm c}$.
By detecting the change in $f_{\rm c}$, we can distinguish the qubit state, {\it  i.e.}, whether  $\sigma_z$ is 1 or -1.
For a flux qubit, the eigenstates of $\sigma_z$ are the two flux states. $\epsilon$ is the bias provided by an external applied magnetic field,
and $\Delta$ corresponds to the tunneling energy between two flux states.

For the qubit-JBA coupled system,  we carried out calculations similar to those without a qubit shown above.
The qubit readout process is well understood by employing knowledge of the quantum behavior in
the time evolution of the JBA without a qubit that  we have already discussed.  

\begin{figure}[h]
\includegraphics[width=4.0cm]{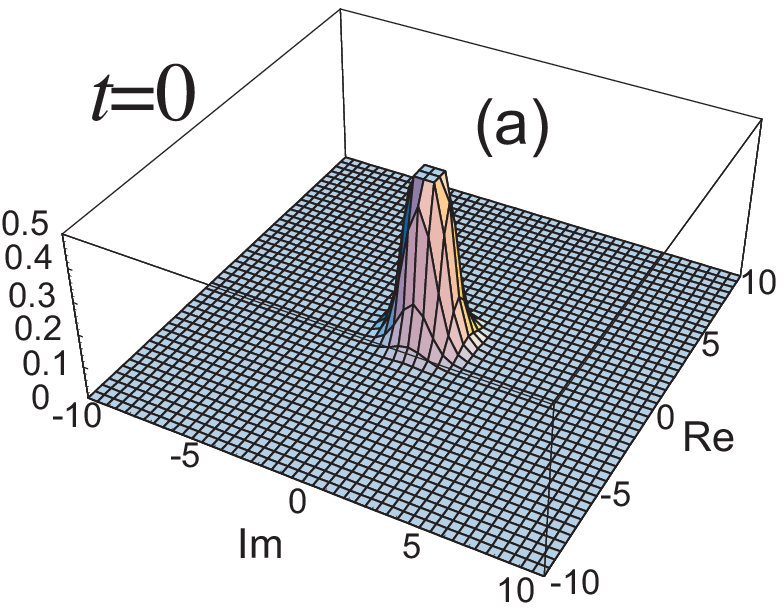}
\includegraphics[width=4.0cm]{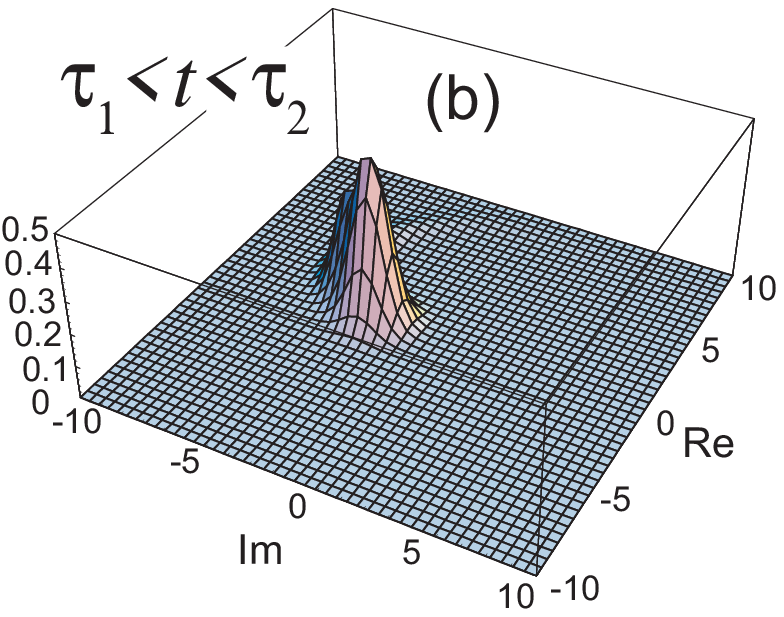}\\
\includegraphics[width=4.0cm]{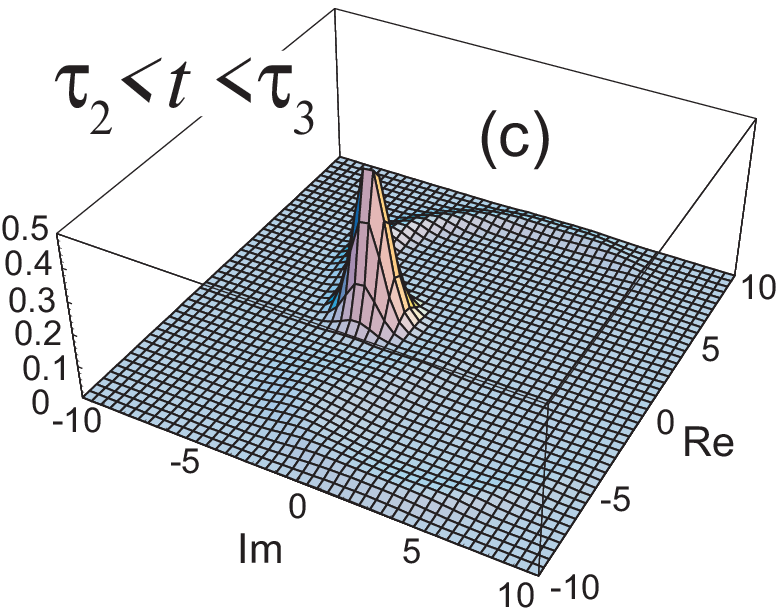}
\includegraphics[width=4.0cm]{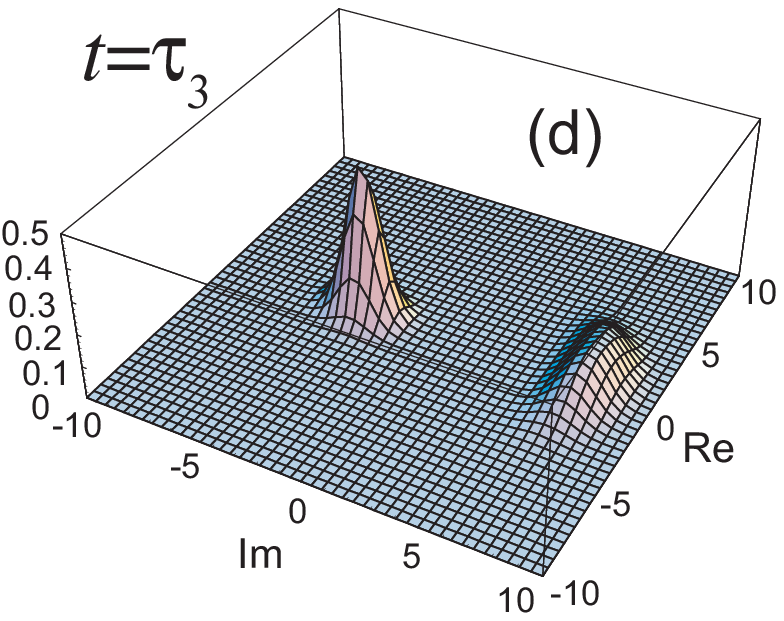}\\
\caption{\label{Dy}
(Color online)  Time-evolution of JBA during  readout. The figures show Q-representations of the JBA oscillator states 
${\rm Tr}_{\rm q}[\rho]$ (in the rotating frame). (a) Beginning of the readout. 
The state of the total system is (schematically) 
$\rho=|\psi_0\rangle\langle \psi_0|$, with $|\psi_0\rangle=1/\sqrt{2}(|g\rangle_q +  |e\rangle_q)\otimes  |G\rangle_{\rm J}$. 
(b) Starting the
transition. Entanglement formation and projection are carried out during this period. (c) During the transition. 
Entanglement has already been  destroyed.
(d) The entire system has become a mixture of classically correlated states.
}
\end{figure}

We show a numerical example of the dynamics during  the qubit readout process in Fig. \ref{Dy}.
JBA parameters are the same as for the above example.
The initial state is a separable state; $(\frac1{\sqrt{2}} | g \rangle_{\rm q}+\frac1{\sqrt{2}} | e \rangle_{\rm q})\otimes |G\rangle_{\rm J}$, that is,
the qubit is in a superposition. Here, $|g\rangle_{\rm q}$ and 
$|e\rangle_{\rm q}$ 
are the ground and excited states of the qubit, respectively.
Qubit parameters are $\epsilon =0.2 \Omega$, $\Delta/\epsilon=1/2$. The coupling between the qubit and the JBA
is set at $k=0.001\Omega$. The driving force $f$ is increased from 0 to $0.012 \Omega$ ( slightly larger than $f_{\rm c}$
of the JBA)  and maintained. This parameter set gives a typical behavior of successful qubit readout. 

The $Q$-representations of the JBA state ${\rm Tr}_{\rm q}[\rho]$ are shown in Fig. \ref{Dy},
where $\rho$ is the density operator of the qubit-JBA coupled system, and ${\rm Tr}_{\rm q}[\cdots]$ 
denotes taking partial trace about  qubit degrees of freedom. 
In  the readout we can distinguish two peaks appearing in Fig. \ref{Dy}(d),
which is the final stage of the readout.
These peaks constitute an incoherent mixture, so they correspond to two possibilities in the measurement result.

To discuss the entanglement between the JBA and the qubit,
we adopt 
$E\equiv {\rm Tr}\left[\rho^2\right]-{\rm Tr}\left[\left({\rm Tr}_{\rm q}[\rho]\right)^2\right]$, 
as a measure of the entanglement.
The reduction in ${\rm Tr}\left[\left({\rm Tr}_{\rm q}[\rho]\right)^2\right]$ is the purity decrease in the reduced
density operator of the JBA, that contains the decrease due to both decoherence and the entanglement 
formation.
The reduction in  ${\rm Tr}\left[\rho^2\right]$ of the total system 
corresponds to the decrease due to decoherence. Therefore, $E$ defined above 
shows the
strength of entanglement. 

\begin{figure}[h]
\includegraphics[width=5.5cm]{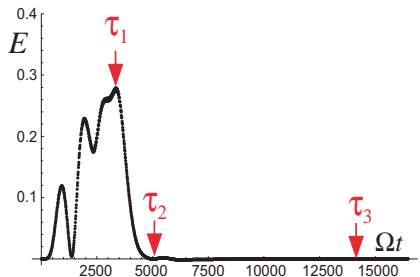}
\caption{\label{fig:entan} (Color on line) Time variation of the entanglement between the JBA and the qubit.  
$\tau_i (i=1,2,3)$ corresponds to those in Eq. \ref{eq:en}, that is , $\tau_1$: entanglement formation,
$\tau_2$: projection, $\tau_3$: end of the readout.}
\end{figure}

The time variation of the entanglement measure $E$ is shown in Fig. \ref{fig:entan}. 
This process can be schematically expressed as
\begin{small}
\begin{eqnarray}\label{eq:en}
&\rho(0)=|G \rangle_{\rm JJ}\langle G |\otimes
 \left(\frac1{\sqrt{2}} | g \rangle_{\rm q}+\frac1{\sqrt{2}} | e \rangle_{\rm q}\right)\left( \frac1{\sqrt{2}}_{\rm q}\langle g | +\frac1{\sqrt{2}}_{\rm q}\langle e | \right)\nonumber\\
\rightarrow &\rho(\tau_1)=
\frac1{2}\left(
| G \rangle_{\rm J}| e \rangle_{\rm q}
+|G'\rangle_{\rm J} | g \rangle_{\rm q}\right)
\left(
_{\rm J} \langle G |_{\rm q}\langle e |+_{\rm J}\langle G'|_{\rm q}\langle g |\right)\nonumber\\
\rightarrow & \rho(\tau_2)=
\frac1{2}| G\rangle_{\rm J}| e \rangle_{\rm qJ}\langle G |_{\rm q}\langle e |+\frac1{2}| G' \rangle_{\rm J}| g \rangle_{\rm qJ}\langle G' |_{\rm q}\langle g |\nonumber\\
\rightarrow& \rho(\tau_3)=
\frac1{2}| G\rangle_{\rm J}| e \rangle_{\rm qJ}\langle G |_{\rm q}\langle e |+\frac1{2}| E \rangle_{\rm J}| g \rangle_{\rm qJ}\langle E |_{\rm q}\langle g |.
\end{eqnarray} 
\end{small}
Entanglement  formation and ``projection'' correspond to  the second ($t=\tau_1$) and  third ($t=\tau_2$)  lines of Eq. (\ref{eq:en}), 
respectively. At this moment ($\tau_2$), however, it is impossible to obtain any information about
the qubit from the observed probe (JBA) state because $|G'\rangle$  closely resembles $|G\rangle$ in a  
classical mechanical sense (Fig. \ref{Dy}(b))
although quantum mechanically $_{\rm J}\langle G'|G\rangle_{\rm J} \sim 0$, namely, these two states are orthogonal.
When we increase the driving force, one JBA state $|G'\rangle_{\rm J}$ moves to $|E\rangle_{\rm J}$.
In contrast, the other $|G\rangle_{\rm J}$ does not move significantly. (see, Figs. \ref{Dy}(c),(d))
\  Then ($\tau_3$), we can easily distinguish 
$|E\rangle_{\rm J}$ or $|G\rangle_{\rm J}$. This leads to a good {\em postulation} of 
the qubit state $|g\rangle_{\rm q}$ or $|e\rangle_{\rm q}$,
which brings us to the end of 
the readout.

The measure $E$ is sufficiently quantitative for us to 
discuss the time variation of the entanglement but it does not show the absolute strength of the entanglement.
To estimate the absolute strength we can calculate the {\em entanglement of formation} 
for every eigenstate consisting the total system density operator $\rho(t)$.
For example, the time variation of the value of the most dominant eigenstate is quantitatively proportional to the behavior of $E$. 
However, it almost becomes unity when it reaches its maximum.
The values for less dominant states also almost reaches unity. This means that the correlation between the qubit and 
JBA state becomes almost perfect via the interaction between them.
As a result an ideal JBA readout exhibits 100\% visibility if the qubit relaxation discussed below is negligible.



The backaction on the qubit caused by the measurement is induced as a result of the non-commutation relation between the qubit Hamiltonian and the
interaction Hamiltonian. 
 When the qubit gap $\Delta$ is much smaller than other energies, the interaction commutes $H_{\rm q}$. 
Therefore, the JBA readout
causes only {\em pure dephasing} on the qubit. This does not pollute the measurement result because the measurement itself requires 
the projection onto the $\sigma_z$ basis. This is simply the condition of the ``non-demolition measurement''.
However, when $\Delta$ is not negligible compared with $\epsilon$,  
the measurement simultaneously causes  {\em qubit relaxation}. 
The non-commuting part induces coherent transition between $|g\rangle_{\rm q}$  and $|e\rangle_{\rm q}$  
in the qubit. This coherent transition itself is not hurmful, but when such a transition is accompanied by 
decoherence (linear loss), stochastic energy relaxation in the qubit accumulate and a finite error remains.
In fact, in the numerical example shown above, the average $\langle\sigma_z\rangle$ of the 
qubit deviates slightly (0.1\%)  from the initial value 0 because of  qubit relaxation. Stronger decoherence causes larger deformation 
in the readout result although it is often much smaller than the deformation caused by other factors
not discussed here, such as qubit relaxation as a reslut of decoherence directly attacking
the qubit, even if we use $10^2$ times strong decoherence  of JBA  as in actual experiments.




As described above, the state of the total system is already
divided into separable states $|e \rangle_{\rm q}|G \rangle_{\rm J}$ and $|g \rangle_{\rm q}|G' \rangle_{\rm J}$ 
just after the transition $|G'\rangle_{\rm J} \rightarrow |E \rangle_{\rm J}$
starts. Therefore, ``projection''  itself is successful 
even if the transition takes much longer  in the absence of sufficiently strong decoherence (here, linear loss $\Gamma$). 
However, we cannot
distinguish $|G \rangle_{\rm J}$ and $|E \rangle_{\rm J}$ until the transition finishes. Then, the readout fails unless
we can wait and maintain the JBA state until the transition is complete.

In summary,  we analyzed the quantum dynamics of the density operator of a system composed of a qubit and a JBA 
as the probe of the qubit state readout.
From the analysis results, we have succeeded in extracting the  essential feature of  the JBA readout process 
of a superconducting qubit. 

We thank M. Devoret and H. Mabuchi for fruitful discussions.
This work received support from KAKENHI(18001002, 18201018).

\end{document}